\documentclass[journal=jacsat,manuscript=article]{achemso}

\usepackage[version=3]{mhchem} 
\usepackage{lipsum} 
\usepackage{color}
\usepackage{graphicx}
\usepackage{dcolumn}
\usepackage{bm}
\usepackage{bbold}
\usepackage{upgreek}
\usepackage{amsfonts,color}
\usepackage{amsmath}
\author{Hanchen Wang}
\email{hanchen.wang@mat.ethz.ch}
\affiliation{%
Laboratory for Magnetism and Interface Physics, Department of Materials, ETH Zurich, Zurich 8093, Switzerland
}%
\author{Laura van Schie}
\affiliation{%
Laboratory for Magnetism and Interface Physics, Department of Materials, ETH Zurich, Zurich 8093, Switzerland
}%
\alsoaffiliation{%
Department of Physics, ETH Zurich, Zurich 8093, Switzerland
}%
\author{Adam Erickson}
\affiliation{%
Department of Physics, ETH Zurich, Zurich 8093, Switzerland
}%
\author{Lauren J. Riddiford}
\affiliation{%
Laboratory for Mesoscopic Systems, Department of Materials, ETH Zurich, 8093, Zurich, Switzerland
}%
\alsoaffiliation{%
PSI Center for Neutron and Muon Sciences, 5232 Villigen PSI, Switzerland
}%

\author{Davit Petrosyan}
\affiliation{%
Laboratory for Magnetism and Interface Physics, Department of Materials, ETH Zurich, Zurich 8093, Switzerland
}%

\author{Christian L. Degen}
\affiliation{%
Department of Physics, ETH Zurich, Zurich 8093, Switzerland
}%
\author{Richard Schlitz}
\affiliation{%
Department of Physics, University of Konstanz, Konstanz 78457, Germany
}%
\author{William Legrand}
\email{william.legrand@neel.cnrs.fr}
\affiliation{%
Universit\'e Grenoble Alpes, CNRS, Institut N\'eel, Grenoble 38042, France
}%
\author{Pietro Gambardella}
\email{pietro.gambardella@mat.ethz.ch}
\affiliation{%
Laboratory for Magnetism and Interface Physics, Department of Materials, ETH Zurich, Zurich 8093, Switzerland
}%

\title[An \textsf{achemso} demo]
{Coherent Microwave Driving of Domain Wall Depinning in a Ferrimagnetic Garnet}

\abbreviations{IR,NMR,UV}

\begin{document}

\begin{tocentry}

\hspace{-0.5cm}\includegraphics[width=120mm]{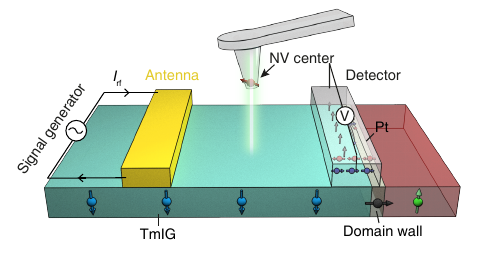}

\end{tocentry}
\clearpage
\begin{abstract}
Coherent control of domain wall dynamics offers a route to fast manipulation of magnetic textures beyond thermally activated motion. We demonstrate resonant excitation of linear and nonlinear dynamics of a pinned domain wall in a ferrimagnetic garnet thin film driven by a microwave field. Using scanning nitrogen-vacancy magnetometry and nonlocal spin-pumping measurements, we identify a low-frequency mode inside the magnon gap, originating from the localized oscillatory motion of a domain wall across a pinning line defined by a Pt stripline. Upon increasing the microwave drive into the nonlinear regime, this mode enables domain wall depinning at reduced external magnetic fields. Micromagnetic simulations reveal a progression from localized oscillations to partial relocation between pinning sites and, ultimately, complete escape from the pinning region with increasing driving power. These results establish resonant excitation of domain walls at engineered pinning sites as a mechanism for manipulating magnetic textures via localized nonlinear dynamics.

\end{abstract}


\newpage
The dynamics of magnetic domain walls (DWs) is central to magnetization reversal processes~\cite{hubert1998magnetic}. In thin films with perpendicular anisotropy (PMA), the nucleation, depinning, and propagation of DWs by an applied magnetic field~\cite{kirilyuk1997magnetization,ferre2013universal} or current~\cite{fukami2011current,emori2013current,yang2015domain,manchon2019current} determine the functionality of magnetic storage~\cite{blasing2020magnetic,franken2012shift,raymenants2021nanoscale} and logic devices~\cite{Luo2020Current,alamdar2021domain,venkat2024magnetic}. Furthermore, DWs can be used to generate and phase-shift spin waves~\cite{garcia2015narrow,hollander2018magnetic,hertel2004domain} or as reconfigurable spin wave channels~\cite{wagner2016magnetic,hartmann2021nonlocal}, which makes them very attractive for applications in magnonics~\cite{yu2021magnetic,petti2022review}. By selectively exciting the internal eigenmodes of DWs, their dynamics can be singled out from the surrounding domain magnetization and actively controlled~\cite{wagner2016magnetic,chen2025observation,zhang2025switchable,liu2019current,albisetti2018nanoscale,camara2017magnetization}. Ferrimagnetic iron garnet thin films with PMA offer an ideal platform to study these phenomena owing to their low damping~\cite{prestwood2025spin,wang2025ultrathin,ding2020nanometer}, high DW mobility~\cite{caretta2020relativistic,velez2019high}, and compatibility with coherent magnon transport~\cite{evelt2018emission,hamalainen2018control,fan2023coherent,han2019mutual}. 

A central challenge in this context is the controlled depinning of DWs from nanoscale energy barriers, which defines the threshold for their mobility and determines device operability. DW pinning and depinning have been extensively investigated in metallic ferromagnets and ferrimagnets using field- and current-induced torques~\cite{chauve2000creep,koyama2011observation,haazen2013domain,gorchon2014pinning,nguyen2014elementary,diazpardo2017universal,woo2017magnetic,caballero2017excess,jeudy2018pinning,diazpardo2019common,gehanne2020strength}, but remain comparatively less explored in insulating magnets~\cite{novoselov2002domainwall,jiang2013mapping,velez2022skyrmion,li2024magnetic,jeudy2025crossover}. In these materials, the low damping and the possibility of accessing DW-specific eigenmodes offer a unique pathway for selective, low-power control of localized magnetic textures.

In this work, we demonstrate coherent nonlinear driving and depinning of DWs in a ferrimagnetic insulator. We deposit a 600-nm-wide Pt stripe on a Tm$_3$Fe$_5$O$_{12}$ (TmIG) thin film to create an artificial pinning line by modification of the local magnetic anisotropy field underneath Pt~\cite{lee2023large,lee2020interfacial,wang2025current}. We excite the DW by a microwave magnetic field and combine scanning NV magnetometry~\cite{hong2013nanoscale,rondin2014nv} and nonlocal spin pumping~\cite{saitoh2006conversion,cheng2020nonlocal,wang2024broad} to directly identify a distinct resonance mode associated with the pinned DW. Micromagnetic simulations confirm that this mode corresponds to a localized oscillatory DW motion. When the excitation frequency matches the pinned-DW resonance, nonlinear coherent driving lowers the depinning threshold, enabling efficient release of the DW. These mechanisms enable resonant control of DWs within ferrimagnetic insulator thin films. 
Our results establish resonant depinning of domain walls as a mechanism for frequency-selective and low-threshold manipulation of magnetic textures in ferrimagnetic insulators.

\begin{figure}
\hspace{-0.5cm}\includegraphics[width=120mm]{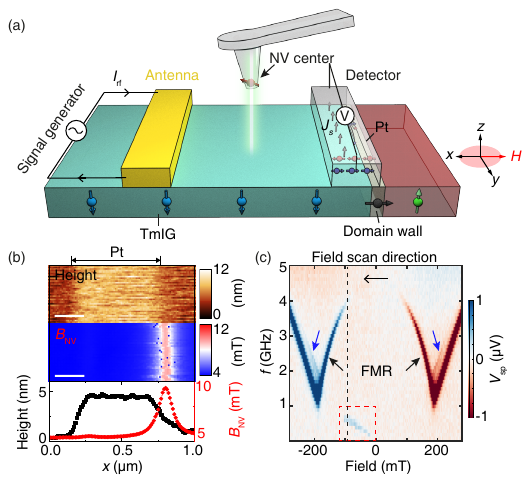}
\caption{(a) Schematic of the pinned DW probed by NV magnetometry, microwave antenna, and nonlocal spin pumping detection. (b) Topology and stray-field maps of pinned DWs imaged by scanning NV magnetometry under the Pt nanostripe with a bias field of 5~mT applied along the $x$ axis. Linecuts of height (black) and stray field (red) show DW pinning at the Pt edge. Scale bar is 200~nm.
(c) Nonlocal spin pumping spectra measured in the Pt nanostripe, showing domain and DW modes (red dashed window) as a function of magnetic field applied along $x$. The microwave power is +5~dBm.}
\label{fig1}
\end{figure}

For our investigation, we employed a 12-nm-thick epitaxial thin film of TmIG, grown on a (111)-oriented yttrium scandium gallium garnet (YSGG) substrate using high-temperature radio-frequency (rf) magnetron sputtering~\cite{wang2025ultrathin,legrand2024lattice}. This film exhibits strong PMA, making it highly suitable for hosting stable magnetic DWs. The Gilbert damping parameter of the uncapped and Pt-capped TmIG film is approximately 0.010 and 0.015, respectively. More details on the sample growth and characterization, including X-ray reflectivity, X-ray diffraction, ferromagnetic resonance (FMR), and magnetization, can be found in Supporting Information (SI)~\cite{Legrand2025}. From the measurements, we obtain a saturation magnetization of $\mu_0M_s \approx 134$~mT, while FMR reveals an effective magnetization $\mu_0M_{\rm eff} = \mu_0M_s - \mu_0H_{\rm ani} \approx -202$~mT. Their difference  indicates a PMA effective field $\mu_0H_{\rm ani} \approx 336$~mT and a PMA energy density of 18~kJ/m$^3$.

Figure~\ref{fig1}(a) illustrates the device used in this study, which includes a Ti(10)/Au(90) nanostripe antenna to inductively excite magnetization dynamics, and a Pt(5) nanostripe positioned $2~\upmu\mathrm{m}$ away (center-to-center) for detecting magnetization dynamics via nonlocal spin-pumping and inverse spin Hall effect (ISHE)~\cite{wang2024broad}. Layer thicknesses (in~nm) are indicated in parentheses. To directly probe with high spatial resolution how and where these nanostructured elements pin DWs within the TmIG, we employed scanning NV center magnetometry. As shown in Fig.~\ref{fig1}(b), concurrent NV-based topography and stray-field maps confirm that the Pt nanostripe can effectively pin a DW at its edge under moderate external fields up to 22~mT (see SI).
In NV magnetometry, uniformly magnetized domains in an ultrathin film generate a vanishing stray field and therefore produce little contrast, whereas sharp magnetization variations at domain walls or sample edges lead to strong signals.
The Pt covering TmIG locally induces additional in-plane anisotropy~\cite{lee2023large,lee2020interfacial,wang2025current}, which is expected to favor DW pinning. This DW can, however, be depinned and further displaced by larger external magnetic fields, as we show below. 
From the height map shown in Fig.~\ref{fig1}(b), the Pt stripe width is about 600~nm, and the antenna width is about 1~$\upmu$m (see SI). Fitting the NV-measured stray field profile at the DW location yields a DW width of approximately 33~$\pm$~7~nm (see SI for details), comparable to previously reported values~\cite{velez2019high}.

Next, we turn to the nonlocal spin-pumping measurements. The microwave antenna is connected to a signal generator that delivers a pulse modulated rf signal with a power of +5~dBm to excite magnetization dynamics. 
All powers reported in the main text refer to the incident microwave power provided by the source. The actual power delivered to the device, accounting for microwave-line losses, is estimated in the SI.
The resulting spin-pumping voltage ($V_{\text{sp}}$) is measured electrically by connecting the Pt detector to a lock-in amplifier. Unless otherwise specified, the external magnetic field is applied in-plane (IP) along the $x$-axis. 
All spin-pumping spectra were acquired via a single field sweep for each excitation frequency or power setting, following initialization by a large field to ensure a well-defined magnetic state for each sweep.
As shown in Fig.~\ref{fig1}(c), three distinct modes are clearly observed. The more visible mode, indicated by black arrows, corresponds to the FMR of the uncapped TmIG film; the second mode, indicated by blue arrows, corresponds to the FMR of the Pt-capped TmIG film, where the Pt capping induces an additional in-plane anisotropy~\cite{lee2023large,lee2020interfacial,wang2025current}. Note that the spin-pumping signal from these two FMR modes becomes weak, and almost invisible, at low fields as the domain magnetization naturally aligns along the out-of-plane (OOP) direction due to PMA. This is because the spin pumping voltage is sensitive only to the IP magnetization component parallel to $x$. Only when the IP magnetic field is large enough for the Zeeman energy to overcome the PMA energy (above about 100~mT), the domain magnetization gradually reorients from OOP to IP. During this transition, the resonance frequency decreases. Once the magnetization is fully aligned with the external IP field (above about 200~mT), the resonance frequency increases with field, following Kittel's dispersion. Finally, in addition to these two FMR modes, we observe a weaker mode at lower frequencies and fields, highlighted by the red dashed box in Fig.~\ref{fig1}(c). This mode emerges in the magnon gap of the bulk domain resonance, suggesting a DW resonance~\cite{wagner2016magnetic,zhang2025switchable,liu2019current,chen2025observation,albisetti2018nanoscale,camara2017magnetization}.

The coexistence of different eigenfrequencies at the same external field, e.g., along the black dashed line in Fig.~\ref{fig1}(c), cannot be attributed to the Tm and Fe sublattices, which remain exchange-locked in this frequency range. Instead, it confirms the presence of a nonuniform magnetic texture, such as the pinned DW observed by NV magnetometry. 
The DW mode also cannot be excited by resonant propagating magnons, as no such states exist at frequencies below the gap. The DW dynamics is instead directly excited by inductive coupling to the antenna radiation.
In addition, the spin-pumping voltage indicates that the DW has a N\'eel-type configuration, consistently with the stray field profile measured by NV magnetometry (see SI for details). This DW configuration can be stabilized by the external magnetic field along $x$, or by the interfacial Dzyaloshinskii-Moriya interaction (DMI)~\cite{Beach2019,Ding2019,velez2019high,Wang2025chiral}.

\begin{figure}
\hspace{-0.5cm}\includegraphics[width=120mm]{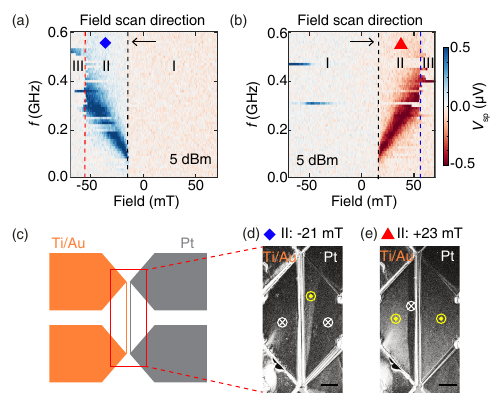}
\caption{(a,b) Nonlocal spin pumping voltage spectra of the DW mode measured while sweeping the magnetic field in opposite directions along $x$ (arrows). The roman numbers indicate the different sweep regions. The applied microwave power is +5~dBm. (c) Schematic of the nonlocal spin pumping device, where the red window delimits the wide-field MOKE imaging area that contains both the antenna and Pt stripe. (d,e) Wide-field MOKE images of Region II while sweeping the magnetic field in opposite directions. The scale bar is 20~$\upmu$m. }
\label{fig2}
\end{figure}

Since in Fig.~\ref{fig1}(c) the DW mode appears only at negative fields during a positive to negative field sweep, we further investigated its dependence on magnetic history. To this end, we conducted high-resolution scans of $V_{\text{sp}}$ at low frequencies while sweeping the magnetic field in opposite directions, as shown in Figs.~\ref{fig2}(a) and~\ref{fig2}(b). The DW mode only appears when the external field is applied opposite to the initial saturation direction, because in this configuration, any small misaligned (within $\pm$~2$^\circ$) OOP field component nucleates a reversed domain that leads to the pinned DW.
The DW mode emerges only after exceeding a threshold opposite field (indicated by black dashed lines), and vanishes when the applied field becomes strong enough to depin the DW from beneath Pt, that is, when the film is fully saturated back into a single-domain state (marked by red or blue dashed lines). For the following discussion, we divide the field sweep into three characteristic regions. (1) Region I: Between the initial saturated state and the DW nucleation threshold; (2) Region II: Field range where the DW remains pinned beneath Pt; (3) Region III: Field strength sufficient to depin the DW. These three regions are also reflected in the background voltages of the spin-pumping spectra in Figs.~\ref{fig2}(a) and~\ref{fig2}(b), which arise from the spin Seebeck effect (SSE) induced by on-chip Joule heating from the antenna~\cite{wang2025orbital}. 

The magnetic hysteresis behavior inferred from the spin pumping spectra is fully reproduced by wide-field MOKE measurements (see SI). Although wide-field MOKE cannot image DWs beneath Pt, it does effectively capture their nucleation and evolution while sweeping the external field after initialization along either direction [Figs.~\ref{fig2}(c-e)].
We conclude that the film is initially saturated into a uniform state (\textit{e.g.}, magnetization pointing down or up) by a large IP field with a small OOP tilt. 
When the field is swept in the opposite direction and crosses the threshold between Region I and II (black dashed line), an ``up" (``down") domain nucleates on the right (left) side of the Pt stripe [Figs.~\ref{fig2}(d) and~\ref{fig2}(e)], with one DW pinned precisely at the Pt boundary. This nucleation requires a finite field to overcome the intrinsic anisotropy barrier, which is present in the entire film but is locally reduced by Pt. 
NV magnetometry and additional hysteresis loops recorded by wide-field MOKE in SI consistently confirm that the resulting DW remains trapped beneath the Pt stripe within the field range where the resonance is observed, and is only released once the external field fully overcomes the pinning potential. 
\begin{figure}
\hspace{-0.5cm}\includegraphics[width=120mm]{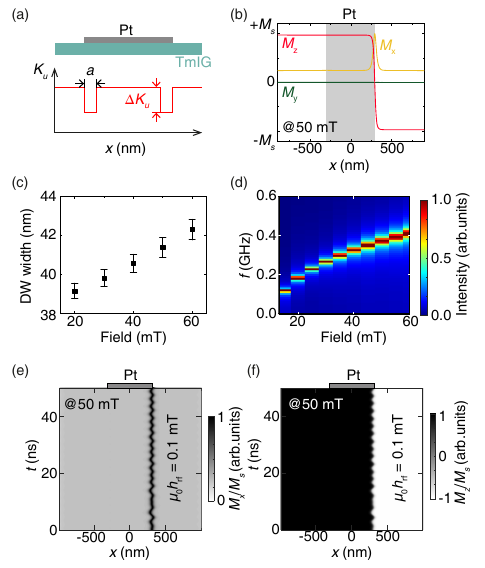}
\caption{(a) Simulation model of reduced PMA ($\Delta K_{\rm u}$) at the Pt nanostripe edges over a width $a$.
(b) Static magnetization profile of a pinned DW at 50~mT, yielding a DW width of $41\pm1$ nm, with in-plane spins aligned along $x$.
(c) Simulated DW width versus external field.
(d) Field-dependent DW resonance spectra obtained from broadband excitation.
(e, f) Time evolution of magnetization components [$M_x$ (e), $M_z$ (f)] under continuous rf excitation at the 50~mT resonance frequency, showing localized oscillatory DW motion at the Pt edge.}
\label{fig3}
\end{figure}

After confirming the hysteretic formation of a pinned DW at the edges of the Pt nanostripe, we investigate its resonance behavior via micromagnetic simulations, using MuMax$^3$~\cite{vansteenkiste2014mumax3}. All magnetic parameters used, such as saturation magnetization, anisotropy, and exchange stiffness, are extracted from SQUID and FMR measurements of the TmIG film. 
To reproduce the observed DW pinning at Pt edges, our simulations consider a PMA reduced by $\Delta K_{\rm u}$ within regions of width $a$ from the edge of the Pt nanostripe [Fig.~\ref{fig3}(a)]. This reduction of PMA is expected from strain or interfacial effects~\cite{lee2023large,lee2020interfacial,wang2025current}. This phenomenological modeling is further supported by additional experiments providing direct evidence for dual DW pinning at each edge of a Pt nanostripe (see SI).
We set $a=10$~nm and $\Delta K_{\rm u}~\approx~-0.5$~kJ/m$^3$, fitted to the 50~mT resonance frequency (see SI). A static up-down DW is initialized at the right edge of a 600 nm-wide Pt nanostripe with a discretization of $3\times3\times3$ nm$^3$ ($xyz$). The external magnetic field is set to 50~mT along the $x$-direction. 

The static magnetization components of the simulated DW are shown in Fig.~\ref{fig3}(b). The simulated DW width is about 41~$\pm$~1~nm, closely matching the one estimated by NV magnetometry (see SI). As evident from the plot, the in-plane magnetization within the DW aligns along the external field direction ($x$), with negligible $y$-component. The DW width increases with field strength as shown in Fig.~\ref{fig3}(c). We analyze the broadband excitation of the DW dynamics in a range of frequencies from 0 to 0.6~GHz and perform a field-dependent simulation of the DW resonance, with the resulting frequency-field relationship shown in Fig.~\ref{fig3}(d). The simulated values match well with the experimental spin-pumping spectra presented in Figs.~\ref{fig2}(a) and~\ref{fig2}(b), lending strong support to the validity of our model.

To visualize the nature of the excitation, we simulate the temporal evolution of the domain-wall dynamics under continuous rf field excitation at the resonance frequency (corresponding to 50 mT). As shown in Figs.~\ref{fig3}(e-f), the DW exhibits an oscillatory motion localized around the pinning region, excited by an rf field in the $z$ direction of 0.1 mT. This relatively small excitation is intentionally chosen to ensure that the system remains well within the linear response regime and to clearly reveal the harmonic motion of the DW.
The $M_x$ and $M_z$ components of the magnetization are plotted as a function of time, clearly illustrating the coherent dynamic nature of the pinned DW resonance. Time-domain data further clarify the mechanism generating the spin-pumping voltages, as detailed in SI. Because the DW oscillation is driven directly by the antenna rf field, and the frequencies correspond to the modes found in micromagnetic simulations, we assign the signal detected below 1~GHz and at +5~dBm to the coherent driving of the DW resonance (see SI).

\begin{figure}
\hspace{-0.5cm}\includegraphics[width=120mm]{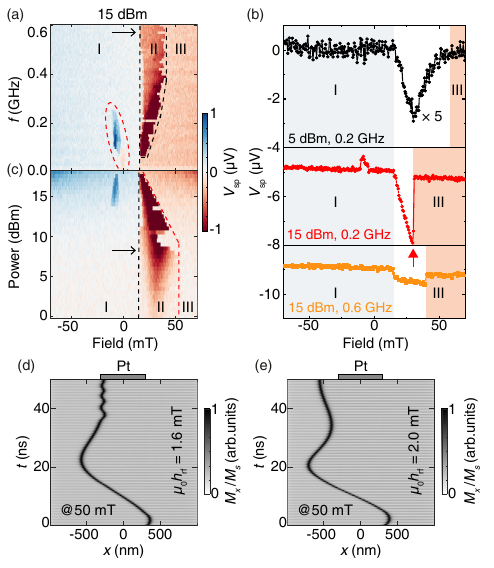}
\caption{
(a) Nonlocal spin-pumping voltage spectra of DW modes measured under a microwave excitation power of $+15~\mathrm{dBm}$. Arrows indicate the field-sweep direction, and dashed lines separate distinct field regions. 
(b) Spin-pumping voltage linecuts at selected excitation powers and frequencies, illustrating resonant-assisted DW depinning.
(c) Power-dependent nonlocal spin-pumping voltage spectra measured at a fixed excitation frequency of $0.2~\mathrm{GHz}$.
(d,e) Temporal evolution of the DW dynamics at resonance, simulated under an excitation field of $1.6~\mathrm{mT}$ and $2.0~\mathrm{mT}$ at a bias field of $50~\mathrm{mT}$, respectively. The $M_x$ magnetization component is shown.
}
\label{fig4}
\end{figure}

We next explore the nonlinear excitation regime. Fig.~\ref{fig4}(a) shows field-dependent spin-pumping measurements at a high microwave power of $+15~\mathrm{dBm}$. Compared with the low-power spectra presented in Fig.~\ref{fig2}(b), several nonlinear features emerge. The first observation is that the field range corresponding to Region~II is no longer identical at different excitation frequencies. The depinning field required to reach the single-domain state (Region~III) is markedly reduced at frequencies below $0.4~\mathrm{GHz}$, coinciding with the DW oscillatory mode. In contrast, for excitation frequencies above $0.4~\mathrm{GHz}$, the depinning field remains essentially constant. 
Consequently, resonant excitation of the DW mode opens an additional depinning channel in external fields that are continuously tunable and significantly lower than the static depinning field.

To more clearly illustrate the resonance-assisted DW depinning, we extracted field-swept spin-pumping voltage linecuts at 0.2~GHz for excitation powers of +5 and +15~dBm, as shown in Fig.~\ref{fig4}(b). At +5~dBm, the amplitude of the DW spatial oscillation is insufficient to trigger depinning, and a nearly symmetric Lorentzian-shaped resonance curve is observed (black dots). In contrast, at +15~dBm, the signal abruptly disappears at the resonant field of 30~mT (red dots and arrow), indicating that the DW escaped from the pinned state. The linecut at +15~dBm and 0.6~GHz (orange dots) shows that the DW remains pinned when excited above the resonance frequency. In this case, the spin pumping voltage originates from the SSE due to thermal magnons generated beneath Pt, and the step-like increase and decrease of the signal reflect external-field-driven nucleation and depinning of the DW, respectively. We further note that the off-resonance threshold field required to reach the single-domain state (Region~III) is power-dependent, as on-chip Joule heating reduces the PMA of TmIG, thereby lowering the external field needed to induce DW depinning. Interestingly, in addition to assisting DW depinning at high power, nonlinear excitations can also facilitate DW nucleation. The positive spin pumping signal in Region I of Fig.~\ref{fig4}(a) (dashed circle) reveals the nucleation of DWs beneath the Pt detector even before any field-driven domain nucleation occurs. However, the mechanism leading to DW nucleation in this case is unknown.

To better illustrate experimentally the role of microwave power in DW depinning, we fixed the excitation frequency at 0.2~GHz and conducted power-dependent measurements, of which the resulting spectra are shown in Fig.~\ref{fig4}(c). We confirmed that the observed depinning behavior is independent of the field sweep direction and the corresponding spectra are provided in SI.
For powers up to +9~dBm, the field range over which the DW stays pinned remains nearly unchanged. However, once the microwave power exceeds this threshold, coherent resonant driving of the DW mode assists the depinning. With further increasing power, the depinning field decreases almost linearly with power, as highlighted by the red dashed lines in Fig.~\ref{fig4}(c). 
A more detailed analysis of the power-dependent transition of the DW mode, from the linear to the nonlinear regime, and a discussion of its connection to the depinning threshold~\cite{lemerle1998domain,jeudy2016universal}, are provided in the SI.
The depinning of the DWs under coherent microwave driving is also reproduced in the simulations. 
When the excitation is close to the energy scale of the pinning potential, the DW depins within nanoseconds, and may be trapped again at the opposite edge of the Pt nanowire [Fig.~\ref{fig4}(d)]. However, with sufficiently strong excitation, the DW depins completely from the Pt nanostripe and moves outside the Pt region [Fig.~\ref{fig4}(e)]. The rf field amplitudes used in Figs.~\ref{fig4}(d) and~\ref{fig4}(e) are chosen to quantitatively match the microwave powers applied in the experiments (see comparison with the experimentally used power in SI)~\cite{fouineau2018semi}.

In summary, we have demonstrated that DWs in insulating ferrimagnetic thin films can be robustly pinned by Pt nanostripes and driven into coherent oscillatory motion by microwave excitation. NV magnetometry and nonlocal spin pumping measurements reveal a distinct resonance mode associated with the pinned DWs, in excellent agreement with micromagnetic simulations. Crucially, when the excitation matches the DW resonance, nonlinear coherent driving provides an efficient pathway to assist DW depinning. This effect is expected to be universal in systems where spin textures can be nucleated and stabilized, for example, through pinning or confinement by patterned nonmagnetic or magnetic structures. The coherent oscillatory modes associated with the motion of localized textures are a general feature of such pinned configurations, providing dynamic control of spin textures in magnetic insulators.

In perspective, the coherent manipulation and depinning of DWs we have observed are not limited to a specific material system, but support a general strategy for the dynamic control of spin textures in magnetic systems. The underlying physics is expected to extend to antiferromagnetic systems~\cite{han2023coherentAFM,wu2024octupoleDW}, where stronger exchange interactions shift the characteristic frequencies upwards. In a low-damping regime, the same dynamics may further give rise to self-induced Floquet magnons or frequency-comb-like responses~\cite{yang2025magnoncomb,yang2025directly}, analogous to recent observations in magnetic vortices~\cite{heins2026floquet,heins2026Controlfloquet} but realized here in a planar-wave geometry, highlighting new opportunities for coherent nonlinear magnonics. Beyond fundamental interest, this platform opens a route toward functional domain-wall-based devices in low-damping magnetic insulators, where selective DW excitation, pinning, depinning, and re-pinning processes may be used in low-power and reconfigurable magnonic or spintronic circuits.

\begin{acknowledgement}
We thank Ka Shen and Shangyuan Wang for helpful discussion.
This research was supported by the Swiss National Science Foundation (Grant No. 200021-236524). 
H.W. acknowledges the support of the China Scholarship Council (CSC, Grant No. 202206020091). 
L.J.R. acknowledges support from the ETH Zurich Postdoctoral Fellowship Program (22-2 FEL-006). 
R.S. acknowledges funding by the Deutsche Forschungsgemeinschaft (DFG, Grant No. 425217212). 

\end{acknowledgement}

\begin{suppinfo}

The data underlying this study are openly available in ETH Research Collection~\cite{Data_ETH}. 

The Supporting Information is available free of charge at [link],

Sample growth and characterization, Scanning nitrogen-vacancy microscopy, Magnetic hysteresis and domain evolution from wide-field MOKE measurements,
Mechanism of spin-pumping voltages from pinned DW oscillatory modes, Power calculations of the RF magnetic field used in the simulation, Power-dependent nonlocal spin-pumping voltage spectra obtained with magnetic-field sweeps in the opposite direction, Experimental evidence for dual domain-wall pinning at the edges of a Pt nanostripe, Transition of the DW mode from the linear to the nonlinear regime, and Phase-locked domain-wall dynamics under microwave excitation.

\end{suppinfo}


\end{document}